\documentclass[preprint,12pt]{elsarticle}

\usepackage{amssymb}
\usepackage{amsmath}
\usepackage{float}
\usepackage{csquotes}
\newcommand{\q}[1]{\enquote{#1}}

\journal{the Journal of Manufacturing Processes}

\begin{document}

\begin{frontmatter}



\title{ADDOPT: An Additive Manufacturing Optimal Control Framework Demonstrated in Minimizing Layer-Level Thermal Variance in Electron Beam Powder Bed Fusion}
\author[cmu]{Mikhail Khrenov\corref{cor}}\ead{mkhrenov@cmu.edu}
\author[cmu]{William Frieden Templeton}
\author[cmu]{Sneha Prabha Narra\corref{cor}}\ead{snehanarra@cmu.edu}
\date{May 16, 2024}
\cortext[cor]{Corresponding Author}
\affiliation[cmu]{organization={Department of Mechanical Engineering, Carnegie Mellon University},
            addressline={5000 Forbes Avenue},
            city={Pittsburgh},
            postcode={15213},
            state={Pennsylvania},
            country={United States of America}}

\begin{abstract}
Additive manufacturing (AM) techniques hold promise but face significant challenges in process planning and optimization. The large temporal and spatial variations in temperature that can occur in layer-wise AM lead to thermal excursions, resulting in property variations and defects. These variations cannot always be fully mitigated by simple static parameter search. To address this challenge, we propose a general approach based on modeling AM processes on the part-scale in state-space and framing AM process planning as a numerical optimal control problem. We demonstrate this approach on the problem of minimizing thermal variation in a given layer in the electron beam powder bed fusion (EB-PBF) AM process, and are able to compute globally optimal dynamic process plans. These optimized process plans are then evaluated in simulation, achieving an 87\% and 86\% reduction in cumulative variance compared to random spot melting and a uniform power field respectively, and are further validated in experiment. This one-shot feedforward planning approach expands the capabilities of AM technology by minimizing the need for experimentation and iteration to achieve process optimization. Further, this work opens the possibility for the application of optimal control theory to part-scale optimization and control in AM.
\end{abstract}

\begin{graphicalabstract}
\includegraphics[width=\textwidth]{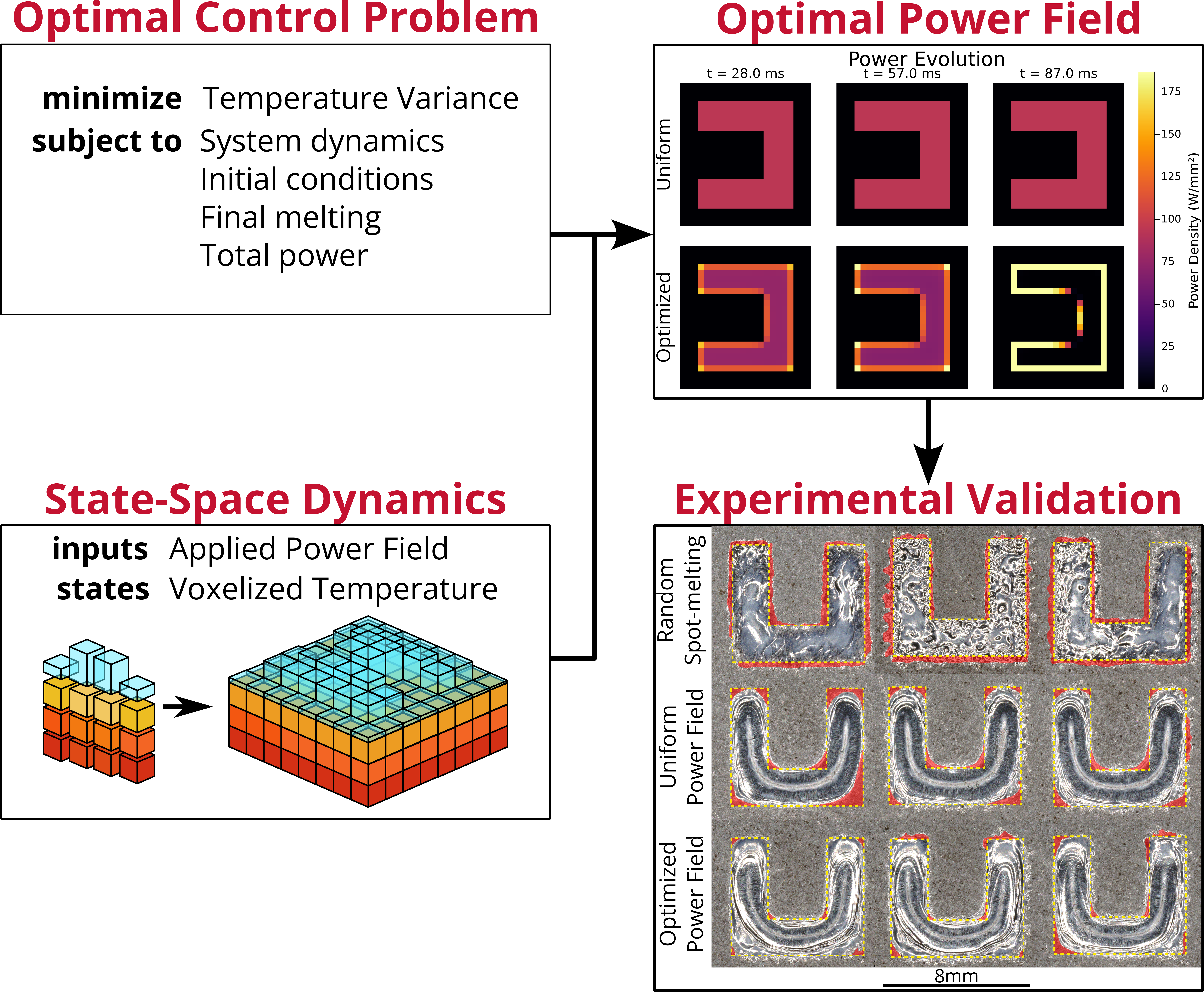}
\end{graphicalabstract}

\begin{highlights}
\item Additive manufacturing processes are modeled at the part-scale in the control theoretic state-space form.
\item Additive manufacturing process planning and optimization for part-scale goals is posed as a general optimal control problem.
\item A software framework implementing this modular optimal control approach is developed.
\item The approach and software framework are applied to generate process plans which minimize cumulative thermal variance across layers in electron beam powder bed fusion.
\item The optimized process plans are shown to be globally optimal and are validated in simulation and experiment.
\end{highlights}


\end{frontmatter}


\section{Introduction}
\label{sec:introduction}
In recent years, metal additive manufacturing (AM) has advanced in producing complex parts with high dimensional accuracy, low defect content, and a minimal need for fixed tooling via sequential layer-wise melting. Currently, process parameters such as beam power and scan velocity are generally held at constant values during the build process, as are the scanning strategies that dictate the beam path. However, the dynamic nature of AM processes means that the use of static parameters and scanning strategies often leads to spatially varying temperature histories both within and across layers. This variation can cause localized over-heating leading to over-melting, stochastic defects, and heterogenous microstructures. Thus, these variations ultimately affect the quality and performance of as-fabricated AM parts \cite{XU2022112065, FosterBeese2017, DEBROY2018112, SCHWALBACH2022143853}.

In current process development approaches, key parameters are evaluated using process maps, where differing combinations of beam power, scanning velocity, and scanning strategy are experimentally evaluated through a grid-based search, and contours of static process parameter combinations that minimize defects are plotted  \cite{ProcessMaps, BeuthPatent, ZHANG2021102018}. While this method of process parameter selection can succeed in avoiding defects for near steady-state conditions, thermal transients during the fabrication process may shift the ideal process parameters for defect avoidance \cite{NARRA2018160, FRIEDENTEMPLETON2024119632}. Due to their static nature, grid-based search process mapping approaches are not suitable for mitigating and exploiting the dynamic nature of metal AM.

\begin{figure}[!htb]
    \centering
    \includegraphics[width=0.5\textwidth]{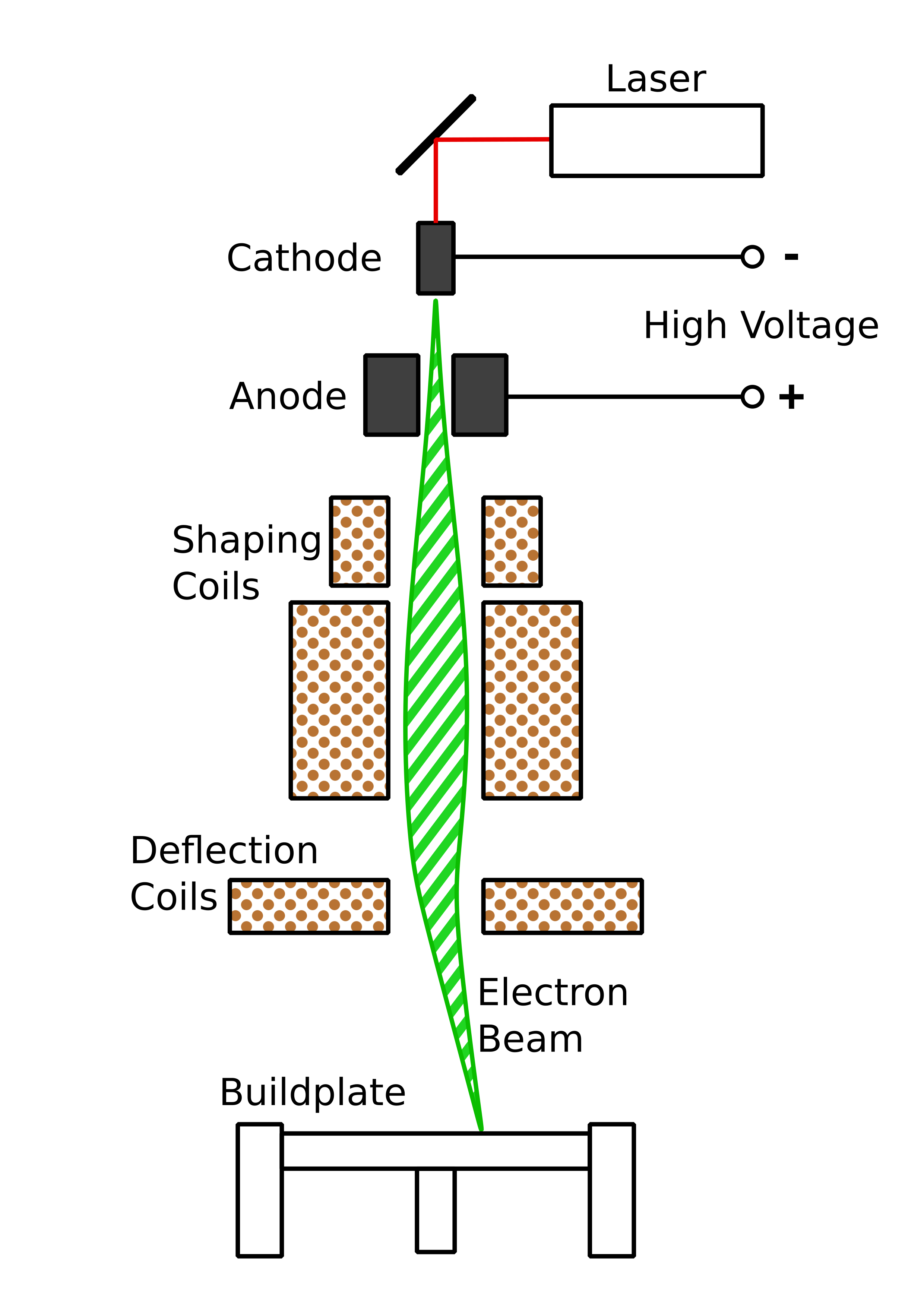}
    \caption{Diagram of electron beam powder bed fusion (EB-PBF).}
    \label{fig:EB-PBF-diagram}
\end{figure}

Just as the limitations of process mapping approaches have become increasingly apparent, new AM hardware has been developed which has rendered grid-based experimental parameter search intractable due to an expansion of processing capabilities. One such case is electron beam powder bed fusion (EB-PBF), shown in Fig. \ref{fig:EB-PBF-diagram}, which melts metal powder using a high power electron beam with powers of up to 6 kW, steered by an electromagnetic field with scanning velocities of up to 4 km/s \cite{freemelt}. Making full use of these capabilities requires highly-complex scanning strategies, such as \enquote{spot melting}, wherein the beam rapidly jumps between points across a layer \cite{babu_spot}. Thus, the process never enters a steady state regime such as a continuous moving meltpool in laser powder bed fusion (L-PBF) AM processes. This in turn rapidly increases the number of decision variables if one were to use grid-search based methods. As a result, rather than exhaustive search, process parameter and scanning strategy decisions are made with limited heuristic methods that do not adequately capture the dynamics of the process. For instance, one approach is random sampling with \enquote{neighbor cooldown}, in which, after a spot has been chosen, its neighbors are excluded from sampling for some number of time steps. While this reduces hot spot formation, it does not directly consider the global temperature evolution \cite{babu_spot} and cannot guarantee global thermal uniformity. Similar challenges and opportunities arise in other recent AM technologies, such as laser area printing \cite{seurat} and large-scale multi-beam systems \cite{vulcanforms}.

One approach to mitigate transient deviations in AM processes has been the application of feedback control to improve the consistency of thermal conditions in the neighborhood of the heat source. This has typically been done by monitoring a proxy quantity, such as meltpool width or bead height, and applying a single input, single output (SISO) controller to stabilize this quantity. Such approaches have shown success in improving geometric consistency and accuracy, both in slower directed energy deposition (DED) type processes (travel speeds in the range of 10 mm/s) \cite{Gibson2020,Li2018,Li2021,Tang2021,Wang2021,Xia2020} and in the L-PBF process \cite{CRAEGHS2010505, Shkoruta2021, WANG2023103449}. However, real-time feedback controllers are challenging to implement on high-speed processes such as EB-PBF. While others have used model-based feedforward to successfully achieve similar goals in L-PBF, the control goals fundamentally remain local to the melt-pool \cite{WANG2020100985, DRUZGALSKI2020101169, Forslund2021, Xiong2022, Shi2023}. Further, while undoubtedly valuable in improving process consistency, such local controllers are insufficient to control the full-part temperature distribution, and thus ensure uniform outcomes in terms of microstructure, stresses, and properties, as these will not only be affected by thermal conditions experienced when a region is subjected to heat directly, but also from the later cooling and re-heating inherent to melting-based AM \cite{XU2022112065, OLLEAK2024100197}.

In order to address the issue of part-scale thermal control, other work has focused on dynamical modeling and process planning. One method uses a greedy algorithm to select the next feature to be scanned at every time-step by exhaustive search minimizing a thermal non-uniformity metric \cite{smartscan}. This is able to achieve significant improvements in uniformity and demonstrable reductions in residual stress. An alternative method has posed scan ordering as a mixed integer optimization problem, and successfully minimized a linearized thermal deviation metric \cite{Schmidt2023}. However, the algorithm of \cite{smartscan} suffers from the short-sightedness of the greedy search, resulting in transients which must in turn be mitigated by heuristic methods. On the other hand, the mixed integer optimization of \cite{Schmidt2023} is limited by its NP-completeness, and so suffers from exponential time complexity as more features are added \cite{Karp1972}. Further, both methods take a fundamentally discrete approach to planning and do not vary process parameters with time. Despite their limitations, these works still achieve significant improvements at the part scale by altering scanning strategies, and show promise in using physics-based dynamic modeling and mathematical optimization for process planning. As such, there is plenty of opportunity to explore in this direction.

In order to address these challenges, we propose leveraging the tools of optimal control and trajectory planning for use in AM. To enable optimal control of AM systems, we present a general form for modeling melting-based AM processes in state-space, along with analysis of selected system models. We formulate a general optimal control problem applicable to layer-wise fabrication governed by these models, and implement this framework for \textbf{add}itive manufacturing \textbf{opt}imal control (ADDOPT) in software. We then demonstrate the ADDOPT framework in the problem of minimizing thermal variance in high-speed, high-power EB-PBF AM. To this end, we present the specific system model assembled via the framework, demonstrate the convexity of the resulting problem, and compute globally optimal process plans for an example geometry. We then convert these optimal process plans to usable machine instructions and compare their performance to current EB-PBF methods both in simulation and through experiment. Finally, we discuss the future work, applications, and implications of the framework and the optimal control approach to AM.

\section{Modeling Additive Manufacturing in State-Space}
\label{sec:modeling}
The tools of modern control theory, including optimal control, have been developed to work with systems represented by a finite vector of state variables $x$ and input variables $u$, with the system dynamics modeled by a first order ordinary differential equation of $x$ and $u$, of the form
\begin{equation}
    \dot{x} = f(x, u)
    \label{eq:state_space_general}
\end{equation}

In order to enable a general framework for optimizing arbitrary AM processes, where the dominant physical effects, required fidelity, or input capabilities may differ from process to process, we propose that any such process can be decomposed into two primary coupled segments: the transport model ($\dot{s}$)
and the input model ($\dot{r}$).
\begin{equation}        
    \begin{split}
        \dot{s} &= f_a(s) + f_i(r,u)\\
        \dot{r} &= f_r(r, u)
    \end{split}
    \label{eq:am_model_general}
\end{equation}

The transport model describes the underlying transport phenomena inherent to a given AM process's environment. In the case of powder bed fusion, this is primarily energy transport via thermal conduction, alongside convection and radiation, but in the case of directed energy deposition AM processes -- which add material at the site of heating -- will also include bulk mass transport. The distinguishing feature of the transport model is that it is both spatial and temporal, and most of the phenomena concerned are governed by partial differential equations. However, to make the problem amenable to numerical analysis and optimization, these states and dynamics are discretized. As such, in our framework the transport model supplies the transport state variables $s$, and the autonomous dynamics ($f_a$) of $s$ in (\ref{eq:am_model_general}).

Meanwhile, the input model describes the actuation the system is subject to, i.e. a laser beam, electron beam, or welding torch. It provides both how this actuation mechanism affects the transport quantities of the part through a forcing term ($f_i$) as well as any internal state variables and dynamics of the actuator ($r$ and $f_r$ respectively) and its constraints.

By combining the transport and input model states, $s$ and $r$, into a combined state vector, $x$, we may write the combined process dynamics in standard state-space form for use with optimal control techniques. 
\begin{equation}
    x = \begin{bmatrix}
        s\\
        r
    \end{bmatrix},~ 
    \dot{x} = f(x, u) = \begin{bmatrix}
        f_a(s) + f_i(r,u)\\
        f_r(r, u)
    \end{bmatrix}
    \label{eq:am_model_combined}
\end{equation}

\subsection{Example Transport Model -- Voxelized Conduction and Convection}
One transport model widely applicable across AM processes and problems -- especially powder bed fusion -- is voxelized conduction/convection, with some prior work presenting finite differences \cite{smartscan} or finite element formulations \cite{Wood2023Theory, wood2019seeing, wood2021ensemble}. Here we present a finite-volume based formulation.

While the temperature profile in a material is continuous and governed by Fourier's Law,
\begin{equation}
    q = -k \nabla T
    \label{eq:fourier}
\end{equation}
if we discretize the part into $l$ by $l$ by $l$ isothermal voxels and consider their internal energy balance, we can write the $i^{\text{th}}$ voxel's internal energy as $E_i = c_p \rho l^3 T_i$, assuming constant density, $\rho$, and specific heat capacity, $c_p$. 

The voxelized dynamics then arise from applying finite volume conduction between voxels, convection to surroundings, and conduction into the baseplate. This results in the following equations for inter-voxel conduction, baseplate conduction, and convection.
\begin{align}
    \dot{q}_{ij}^k &= \frac{k}{l} (T_i - T_j) A_{ij}\\
    \dot{q}_{i0}^k &= \frac{k}{l} (T_i  -  T_{0}) A_{i0}\\
    \dot{q}_{i\infty}^c &= h (T_i - T_{\infty}) A_{i\infty}
\end{align}
Where $A_{ij}$ is the area shared between two adjacent voxels ($l^2$ for the uniform grid case), $A_{i0}$ is the area a voxel has in contact with the baseplate,  $A_{i\infty}$ is the area a voxel has exposed to the environment, $T_0$ is the baseplate temperature, $T_\infty$ is the ambient temperature , and $h$ is the convective heat transfer coefficient.

\begin{figure}[!htb]
    \centering
    \includegraphics[width=\textwidth]{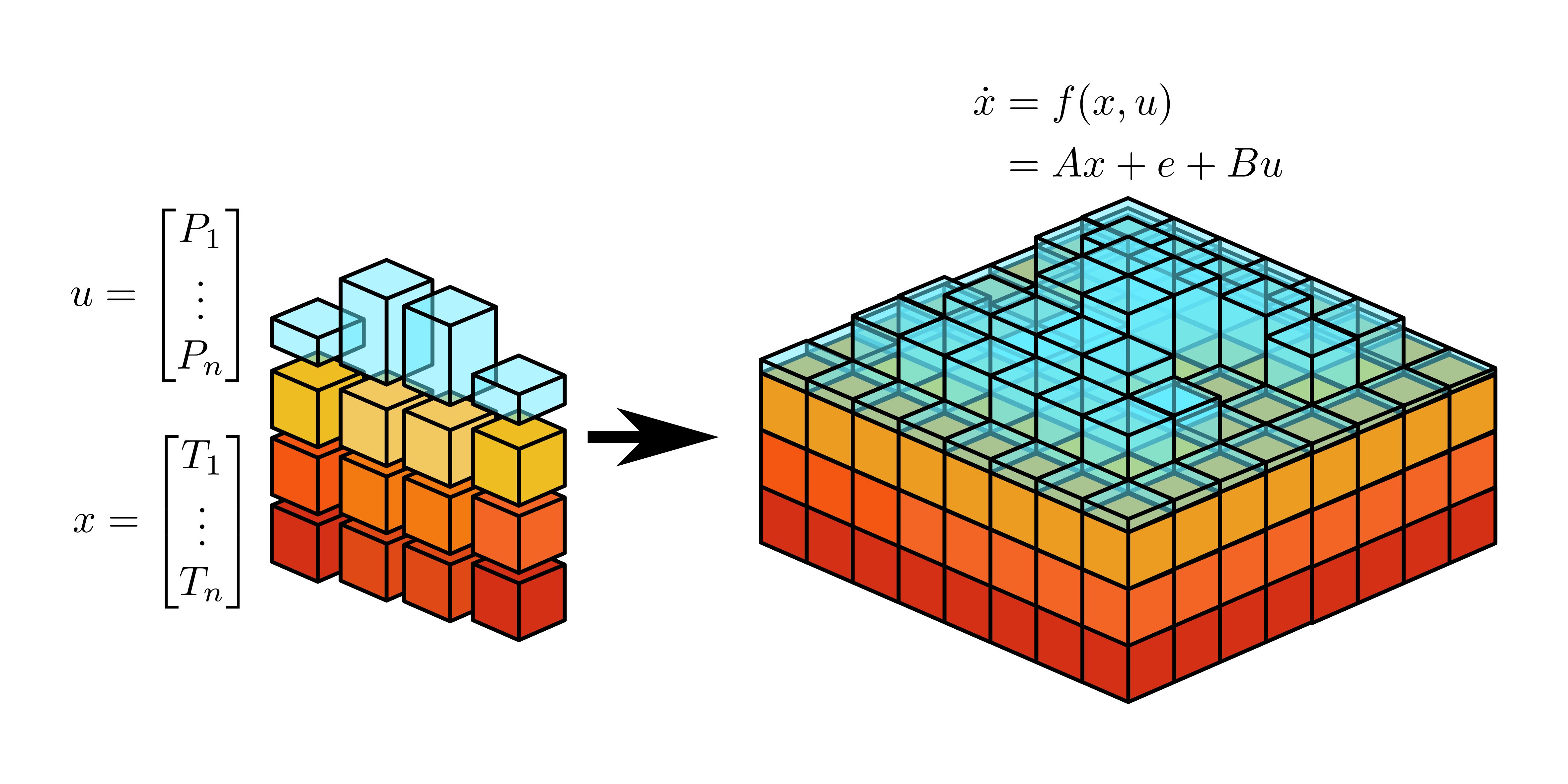}
    \caption{Voxelized conduction transport model with power field input model.}
    \label{fig:voxelized-conduction}
\end{figure}

We can then apply an energy balance to each voxel considering each of these flows from the neighbor set $N_i$ and environment, as well as any external power being added to a voxel, \(P_i\), and reformulate in terms of temperature to get the total per-voxel  temperature dynamics
\begin{equation}
    \dot{T}_i = \frac{-\sum_{j \in N_i} \dot{q}_{ij}^k - \dot{q}_{i0}^k - \dot{q}_{i\infty}^c + P_i}{\rho l^3 c_p}
\end{equation}
If we consider the connections between adjoining voxels to form a graph, combining the dynamics for all voxels as shown in Fig. \ref{fig:voxelized-conduction}, and defining $x = [T_1 \dots T_n]^T$, we obtain the state-space ensemble dynamics of the system 
\begin{equation}
    \begin{split}
        \dot{x} &= -(\frac{\alpha}{l^2}L + \frac{\alpha}{l^4} A_{0} + \frac{h}{C} A_{\infty}) x + (\frac{\alpha}{l^4}A_{0}T_{0} + \frac{h}{C}A_{\infty}T_{\infty})\vec{1} + f_{in}(r, u)
    \end{split}
\end{equation}

Where $L$ is the graph Laplacian of the finite volume mesh, defined by $L = \Delta - \Lambda$ for $\Delta$ degree matrix and $\Lambda$ adjacency matrix, $A_{\infty}$ is the diagonal matrix of $A_{i \infty}$, $A_{0}$ is the diagonal matrix of $A_{i 0}$, $\alpha$ is the thermal diffusivity $k/(\rho c_p)$, and $C$ is the per-voxel heat capacity, $c_p \rho l^3$. Neglecting the power input $f_{in}$, this forms a networked system following consensus dynamics, with elements exposed to the atmosphere or baseplate acting as \q{leaders}, whose characteristics can be analyzed as such \cite{MurrayNetworks}.

Finally, if we define the dynamics matrix $A = -(\frac{\alpha}{l^2}L + \frac{\alpha}{l^4} A_{0} + \frac{h}{C} A_{\infty})$ and the exogenous input vector $e = (\frac{\alpha}{l^4}A_{0}T_{0} + \frac{h}{C}A_{\infty}T_{\infty})\vec{1}$, then we can write the autonomous dynamics in standard linear time invariant form
\begin{equation}
    \dot{x}(t) = A x(t) + e +  f_{in}(r, u)
\end{equation}
This model, while simplified, offers a way to both simulate and optimize full part temperature evolution during AM for given material properties, convection coefficients.

\label{subsec:voxelized_conduction}

\subsection{Example Input Model -- Power Field}
\label{subsec:power_field}
A simple way to model power inputs to a voxelized AM process is to consider the machine as being able to apply a \textit{power field} over the surface of the part/powder. While traditional L-PBF systems with a single laser steered by galvanometer mounted mirrors are not well approximated by this model, as the heat source is localized and relatively slow, technologies are becoming increasingly available that satisfy this model such as EB-PBF and newer L-PBF variants \cite{seurat, vulcanforms, freemelt}. 

The resulting input model can then be stated as
\begin{equation}
    \begin{split}
        f_{in}(r, u) & = Bu\\
        \text{s.t.}~ & u \geq 0\\
                     & \sum_{i=1}^m u_i \leq P_{\max}\\
                     & \sum_{i=1}^m u_i \geq P_{\min}
    \end{split}
    \label{eq:power-field}
\end{equation}
Where $B$ is an $n$ by $m$ matrix mapping the $m$ voxels over which power can be specified to the appropriate elements of the $n$-dimensional state.

\section{Optimal Control Problem}
\label{sec:optimal-control-problem}
Given these models, we can phrase AM process optimization as an optimal control problem. In its most general form, it is a minimization of a cost function ($J$), integrated over every layer-wise build cycle and summed across all $N_c$ build cycles. 
\begin{equation}
\begin{split}
    \min_{x(t), u(t), \{(t_i^c, t_b^c, t_f^c)\}}~& \sum_{c=1}^{N_c} \int_{t_i^c}^{t_b^c} J(t, x(t), u(t)) dt + \int_{t_b^c}^{t_f^c} J(t, x(t), u(t)) dt\\
    \text{s.t}.~& \dot{x}(t) = f(x(t), u(t)) ~\forall t\\
        & x(0) = x_0, ~x(t) \in X ~\forall t\\
        & u(t) \in U ~\forall t \in (t_i^c, t_b^c)\\
        & u(t) = 0 ~\forall t \in (t_b^c, t_f^c)\\
        & t_i^{c+1} = t_f^{c} ~\forall c\\
        & t_i^1 = 0
\end{split}
\label{eq:general_problem_continuous}
\end{equation}
The system is constrained to obey its dynamics ($f$), initial conditions ($x_0$), and the inputs are constrained to remain within the allowable sets ($U$) during the build periods $(t_i^c, t_b^c)$ and to be zero during cooling periods $(t_b^c, t_f^c)$. The state trajectories, input trajectories, and relative lengths of each cycle's build and cooling periods are decision variables.

In the general case where an analytic solution is unavailable, the above is an infinite-dimensional problem in $x(t)$ and $u(t)$. To make the problem amenable to numerical solution, a discretization must be applied. To accomplish this, we make use of the direct collocation approach pioneered in aerospace and robotic applications \cite{DIRCOL}, which transforms the optimal control problem in (\ref{eq:general_problem_continuous}) into a finite-dimensional non-linear program.
\begin{equation}
\begin{split}
    \min_{x_{1:N}, u_{1:N}, \Delta t_{1:N}}~& \sum_{k=1}^N J_k(x_k, u_k, \Delta t_k)\\
    \text{s.t}.~& c(x_k, u_k, x_{k+1}, \Delta t_k) = 0 ~\forall k\\
        & x_1 = x_0\\
        & x_k \in X_k ~\forall k\\
        & u_k \in U_k ~\forall k \in \mathcal{B}\\
        & u_k = 0 ~\forall k \in \mathcal{C}\\
        & \Delta t_k \in (\Delta t_{\min}, \Delta t_{\max})
\end{split}
\label{eq:general_problem_discrete}
\end{equation}
In this approach, the state trajectory is approximated with a sequence of piecewise continuous polynomials and represented by the $N$ \q{knot points}. While a variety of interpolations could be applied, we make use of the cubic spline approach originally proposed in \cite{DIRCOL}. This then results in the dynamic feasibility constraint $\dot{x} = f(x, u)$ being reformulated as a third-order Hermite-Simpson implicit constraint, which we simply write as the constraint $c$, and the objective function being re-written as the sum of its integrations over the $k^{\text{th}}$ timestep. The objective is also discretized ($J_k$).

In order to allow the length of the build and cooling periods to vary, the length of the $k^{\text{th}}$ timestep, $\Delta t_k$, may be allowed to vary, while the transition between build and cooling periods is handled in a hybrid system manner by pre-assigning timesteps to either be in the build set $\mathcal{B}$ or the cooling set $\mathcal{C}$ and applying the input constraints as appropriate.

\section{Implementation}
\label{sec:implementation}
\begin{figure}[!hbt]
    \centering
    \includegraphics[width=0.95\textwidth]{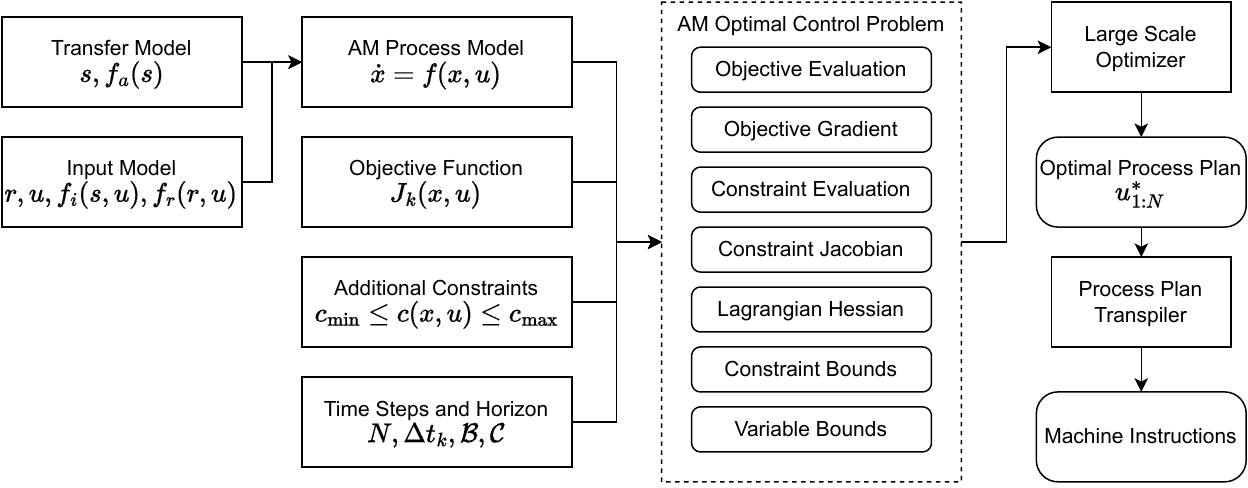}
    \caption{Block diagram of the ADDOPT system.}
    \label{fig:ADDOPT-block-diagram}
\end{figure}

In order to numerically solve the optimization problem presented in Sec. \ref{sec:optimal-control-problem}, we developed a modular system which we refer to as ADDOPT (the \textbf{add}itive manufacturing \textbf{opt}imal control framework), whose overall structure is shown in Figure \ref{fig:ADDOPT-block-diagram}. The system was written in the Julia programming language, implementing the modular system of composing AM process models by combining a transport
and input model with constraints as described in Sec. \ref{sec:modeling}, and then combining the process model with an objective, number of cycles and knot points, and whether time is allowed to vary to fully define the AM optimal control problem. 

The resulting additive problem structure is used to construct a non-linear program (NLP) which is passed to the IPOPT interior point solver \cite{Wächter2006} with the HSL\_MA97 linear solver \cite{HSL_MA97}. Due to the usually high sparsity of the Jacobian of the dynamics in additive processes, ADDOPT automatically computes the sparsity of the constraint Jacobian using the Symbolics.jl package \cite{symbolics_jl} and computes the objective, constraints, objective gradient, constraint Jacobian, and objective Hessian, with sparse forward mode automatic differentiation being used for the constraint Jacobian \cite{RevelsLubinPapamarkou2016, SparseDiffTools}.

The NLP solutions are then marshalled and converted to machine appropriate instructions, with the implementation also providing visualization tools. The source code for ADDOPT and the demonstration on EB-PBF shown in this work is available upon request from the corresponding authors.

\section{Demonstration in Electron Beam Powder Bed Fusion}
\label{sec:demonstration}
We now proceed to demonstrate the practical feasibility and benefits of applying optimal control to metal AM via the ADDOPT framework. As our example case, we consider high power EB-PBF. As previously mentioned, current process optimization is carried out using process maps of fixed power/velocity or power/dwell time combinations for a static scan strategy. However, the underlying process in EB-PBF is fundamentally dynamic and transient. At the same time, the high beam travel speeds possible in EB-PBF open the door to approximating near-arbitrary power fields. As such, both the challenges and opportunities present in EB-PBF make it a good target for applying optimal control to full-part thermal distributions with constraints on the states and inputs.

\begin{figure}[!htb]
    \centering
    \includegraphics[width=0.4\textwidth]{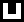}
    \caption{Top-down view of the target geometry for the EB-PBF experiments. White signifies areas to be melted, while black signifies areas to not melt. The varying thicknesses pose a challenge to maintaining uniform thermal distributions for current techniques.}
    \label{fig:target}
\end{figure}

Specifically, we demonstrate how the proposed optimal control approach and software framework allow the generation of process plans which globally minimize the cumulative thermal variance over a target geometry (Fig. \ref{fig:target}) while satisfying spatially variant temperature constraints. 

\subsection{Process Model}
The EB-PBF process model is constructed by combining the voxelized conduction/convection transport model (Sec. \ref{subsec:voxelized_conduction}) with the voxel-wise power field input model (Sec. \ref{subsec:power_field}). This model is appropriate for electron-beam processes due to their extremely fast beam travel speeds, up to 4 km/s in the case of our equipment, as it is possible to move the beam sufficiently fast that one can approximate an arbitrary power field, subject to total power constraints, in time average terms. However, since the cathode power has a slow response time on the equipment used, taking 2 to 5 seconds to reach 3 kW, the beam is brought up to power before printing and is held at constant power while scanning. As a result, both $P_{\min}$ and $P_{\max}$ are set to 3 kW. Notably, as this model is conduction mode, it does not model fluid flow effects, as will be seen in the Results.

The substrate is a circular 316L stainless steel plate with a diameter of 100 mm and a thickness of 10 mm. Experimentally measured material properties are available in the literature. Specifically, the model uses constant properties taken at the solidus temperature \cite{mills}: a thermal conductivity of 31.1 W/m K, a density of 7269 kg/m$^3$, and a specific heat of 720 J/kg K. The part is divided into a uniform grid of voxels with 200 $\mu$m side lengths.

\subsection{Objective}
In order to maximize part quality and uniformity, our goal is to minimize thermal variance over the region to be printed. We will refer to this region as the \q{mask}, or $\mathcal{M}$. Given the discretization of the domain into N voxels, we define the mask vector $\mu$, such that
\begin{equation}
    \mu_i = \begin{cases}
        1 & i \in \mathcal{M}\\
        0 & i \notin \mathcal{M}
    \end{cases}
\end{equation}
With $N_\mu$ being the total number of voxels in the mask, and $M$ being the matrix $M = \text{diag}(\mu)$. We can now consider the thermal variance over voxels in the mask,
\begin{equation}
    \label{eq:thermal-uniformity}
    \begin{split}
l(T) &= \frac{1}{N_\mu} \sum_{i=1}^{N_\mu} (T_i - \bar{T}_\mu)^2 \mu_i\\ 
            &= \frac{1}{N_\mu} (MT - \bar{T}~\mu)^T (MT - \bar{T}~\mu)\\
            &= \frac{1}{N_\mu} T^T (M - \frac{1}{N_\mu} \mu^T\mu)^T (M - \frac{1}{N_\mu} \mu^T\mu) T
    \end{split}
\end{equation}
Let $Q = \frac{2}{N_\mu} (M - \frac{1}{N_\mu} \mu^T\mu)^T (M - \frac{1}{N_\mu} \mu^T\mu)$. Then, the variance cost function can be expressed as the quadratic form
\begin{equation}
    l(T) = \frac{1}{2} T^T Q T
\end{equation}
As can be readily seen by the definition of variance, $Q$ must be positive semi-definite, with zero eigenvalues corresponding to the mean vector (eigenvector in which all the elements included in the mask have the same value) and to eigenvectors corresponding to each element not included in the mask. As such, $\sigma^2(T)$ is a convex function over $T$.

We then seek to minimize the cumulative variance over the print,
\begin{equation}
    \label{eq:cumulative-var}
    J(T(t)) = \int_0^{t_f} \frac{1}{2} T^T Q T dt
\end{equation}
Which we discretize as the summation
\begin{equation}
    J_d(T_{1:N}) = \sum_{k=1}^N \frac{1}{2} T_k^T Q T_k \Delta t
\end{equation}
itself a convex quadratic form.

\subsection{Additional Constraints}
Beyond the constraints on power imposed by the input model, in order to ensure high geometric accuracy, two additional constraints are added on the temperature distribution.

First, to ensure that voxels not in the mask are not melted, an inequality constraint is added limiting the temperature of these voxels to the solidus, $T_s$,
\begin{equation}
    T^i_k \leq T_s ~\forall k \in \{1 \dots N\}, ~\forall i \notin \mathcal{M}
    \label{eq:constr_no_melt}
\end{equation}

Second, to ensure that all voxels that are in the mask are melted, an inequality constraint is added to keep said voxels above the liquidus temperature, $T_l$, at the final build step,
\begin{equation}
    T^i_N \geq T_l ~\forall i \in \mathcal{M}
    \label{eq:constr_full_melt}
\end{equation}

\subsection{Convexity}
Given that the dynamics (and thus dynamics constraints) are linear, the time-steps are fixed, the cost is quadratic with a positive semi-definite $Q$ matrix, and all of the additional equality and inequality constraints are linear, the total optimization problem is a convex quadratic program (QP),
\begin{equation}
\begin{split}    
    \label{eq:eb-ocp}
    \min_{T_{1:N}, u_{1:N}}~& \sum_{k=1}^N \frac{1}{2} T_k^T Q T_k \Delta t\\
    \text{s.t}.~& T_{k+1} = A_d T_k + e_k + B_d u_k ~\forall k\\
        & T_1 = T_0\\
        & T^i_k \leq T_s ~\forall k \in \{1 \dots N\}, ~\forall i \notin \mathcal{M}\\
        & T^i_N \geq T_l ~\forall i \in \mathcal{M}\\
        & \sum_{i} u_k^i = P ~\forall k \in \{1 \dots N\}\\
        & u_k \geq 0 ~\forall k \in \{1 \dots N\}
\end{split}
\end{equation}
As such, it is worth noting that if the problem is feasible, it can be solved to a tolerance $\varepsilon$ with a Newton-based interior point algorithm within $O(\log(1/\varepsilon))$ steps, also known as linear convergence \cite{POTRA2000281}. Thus, applying ADDOPT to this problem will result in an efficient solution for the power trajectory which globally minimizes the cumulative thermal variance.

\subsection{Power Field Approximation}
\label{subsec:approx}
As the electron beam system is not able to directly generate perfect power fields at a given time, a system is needed to approximate the optimal power fields using the high speed beam. 

We can accomplish power field approximation using a greedy model-based algorithm which takes traverse time into account to minimize the squared error between the optimal power field and the time-average achieved power field. The method is as follows:
\begin{enumerate}
    \item Given a time-step with a beam location and prior power history, generate the power field trajectory that would result from traversing to and dwelling at every other voxel
    \item For every voxel, compute the time-average power field that would result from integrating these power fields and the prior power field over some time window
    \item For every voxel, compute the squared norm difference between these time-averaged fields and the optimized field for the current time-step
    \item Select the voxel with minimum squared difference, append its location and the computed power-field to the plan, add the dwell time to the current time, repeat
\end{enumerate}

By explicitly taking the traverse time and power into account, a good approximation of the optimal power field can be achieved.

This algorithm was executed with a model of the EB-PBF machine used in this work, the Freemelt ONE, an open-source machine that allows user-defined beam current, accelerating voltage, spot size, and user-defined scanning strategies. The electron beam has a Gaussian power distribution with a minimum full-width half-maximum spot size of 200 $\mu$m. A 250$\mu$m beam was used in this work.  

The minimum time-step resolution for commands is 1 $\mu$s, with beam motion being modeled as the response of a first order system due to the inductance of the coils. Due to the minimum time resolution, only integer microsecond dwells were allowed in the greedy approximation algorithm, with the motion of the beam being integrated to sub-microsecond resolution.

\section{Results}
\subsection{Numerical Simulation Results}
\begin{figure}[!htb]
    \centering
    \includegraphics[width=0.85\textwidth]{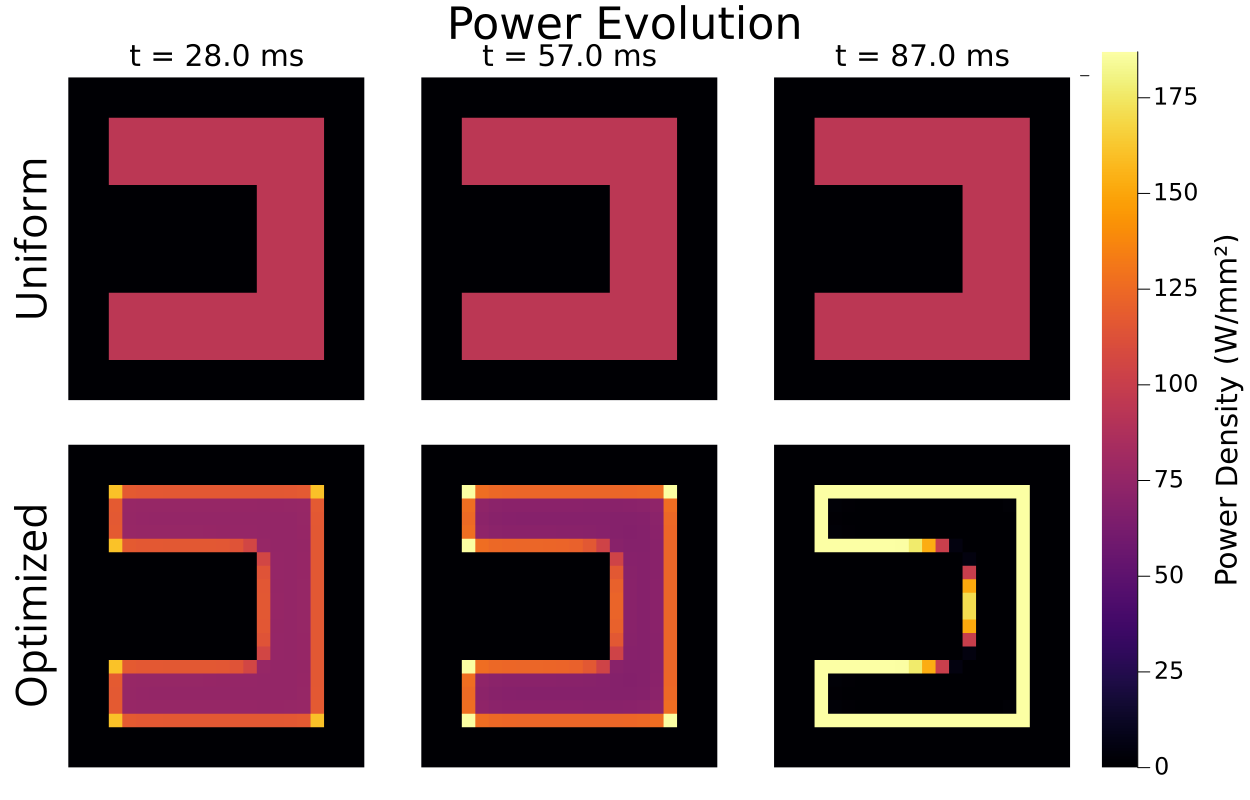}
    \includegraphics[width=0.85\textwidth]{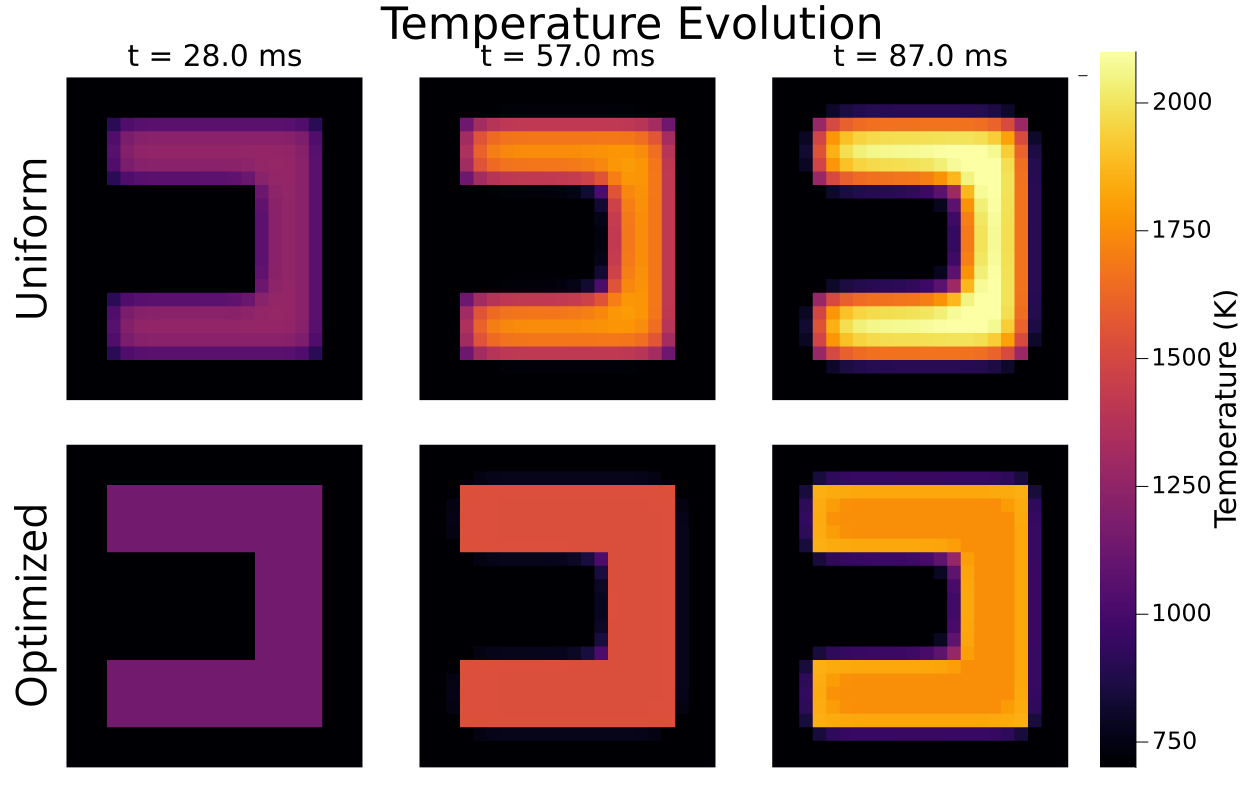}
    \caption{Top: Comparison of the naive power distribution (uniform) and optimized power distribution over the course of the build. Bottom: Comparison of the resulting thermal evolutions. Note the heat build-up in the inner corners of the uniform power field's thermal profile.}
    \label{fig:snapshots}
\end{figure}

We solve the optimization problem to minimize the variance over the geometry shown in Fig. \ref{fig:target} as described in Sec. \ref{sec:implementation}. Solving over 24 by 22 by 4 voxels (2112 voxels total) for 108 timesteps takes 42 minutes using 8 cores of a 4 GHz CPU. We also simulate the temperature evolution resulting from applying a uniform power distribution over the target geometry. A side-by-side comparison of the uniform and optimized power distributions and the resulting temperature distributions is shown in Fig. \ref{fig:snapshots}.

However, as previously mentioned, the electron beam in question can not directly generate the prescribed fields, but must instead approximate them through rapid movement of the beam. We generate approximations using the greedy tracking approach from Sec. \ref{subsec:approx}. These are then also simulated. 

\begin{figure}[!htb]
    \centering
    \includegraphics[width=0.8\textwidth]{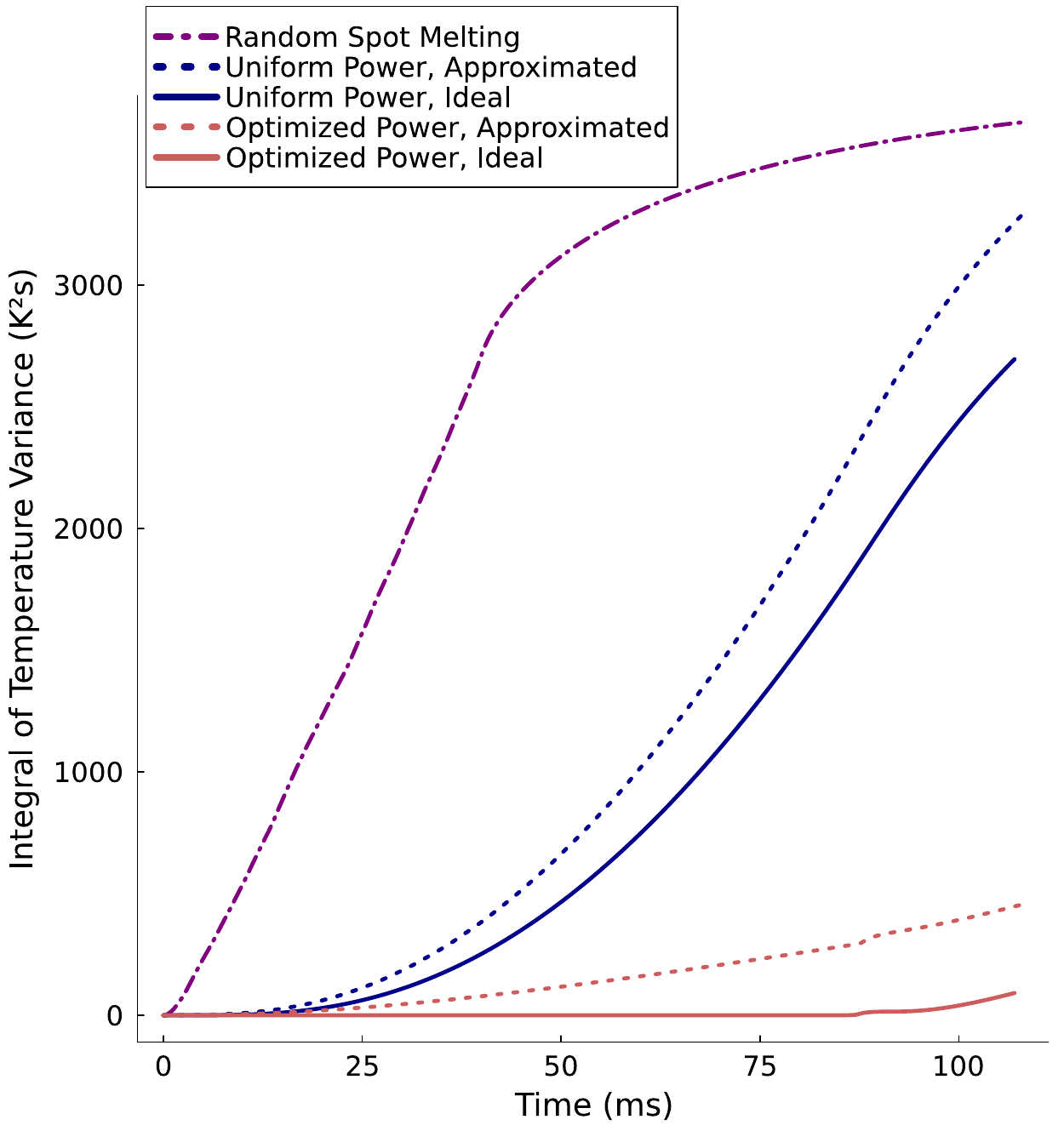}
    \caption{Cumulative thermal variance over the course of the build for random spot melting, an exact uniform power distribution and optimized power distribution, as well as greedy approximations of each.}
    \label{fig:var_comp}
\end{figure}

The resulting cumulative thermal variance for each of the possible build plans is shown in Fig. \ref{fig:var_comp}. As can be seen, the optimized power distribution results in a vastly lower cumulative variance than simply applying a uniform power field. While approximating this field with the greedy tracking approach results in an increase in variance over the ideal power field, the trend remains the same, with 86\% lower cumulative variance than the corresponding uniform field strategy.
Random spot-melting results performs worse than either power field approach, with the highest simulated cumulative temperature variance.

Additional videos visualizing the power and temperature evolution under the various schemes are included in the supplementary materials.

\subsection{Experimental Results}
The random spot melting, uniform power field, and optimized power field beam paths were converted into instruction compatible with the Freemelt ONE EB-PBF system. Nine geometries were printed, with three of each beam path type, on a 316L stainless steel substrate at a power of 3 kW and full-width half-maximum beam spot size of 250 $\mu$m.

\begin{figure}[!htb]
    \centering
    \includegraphics[width=1.0\textwidth]{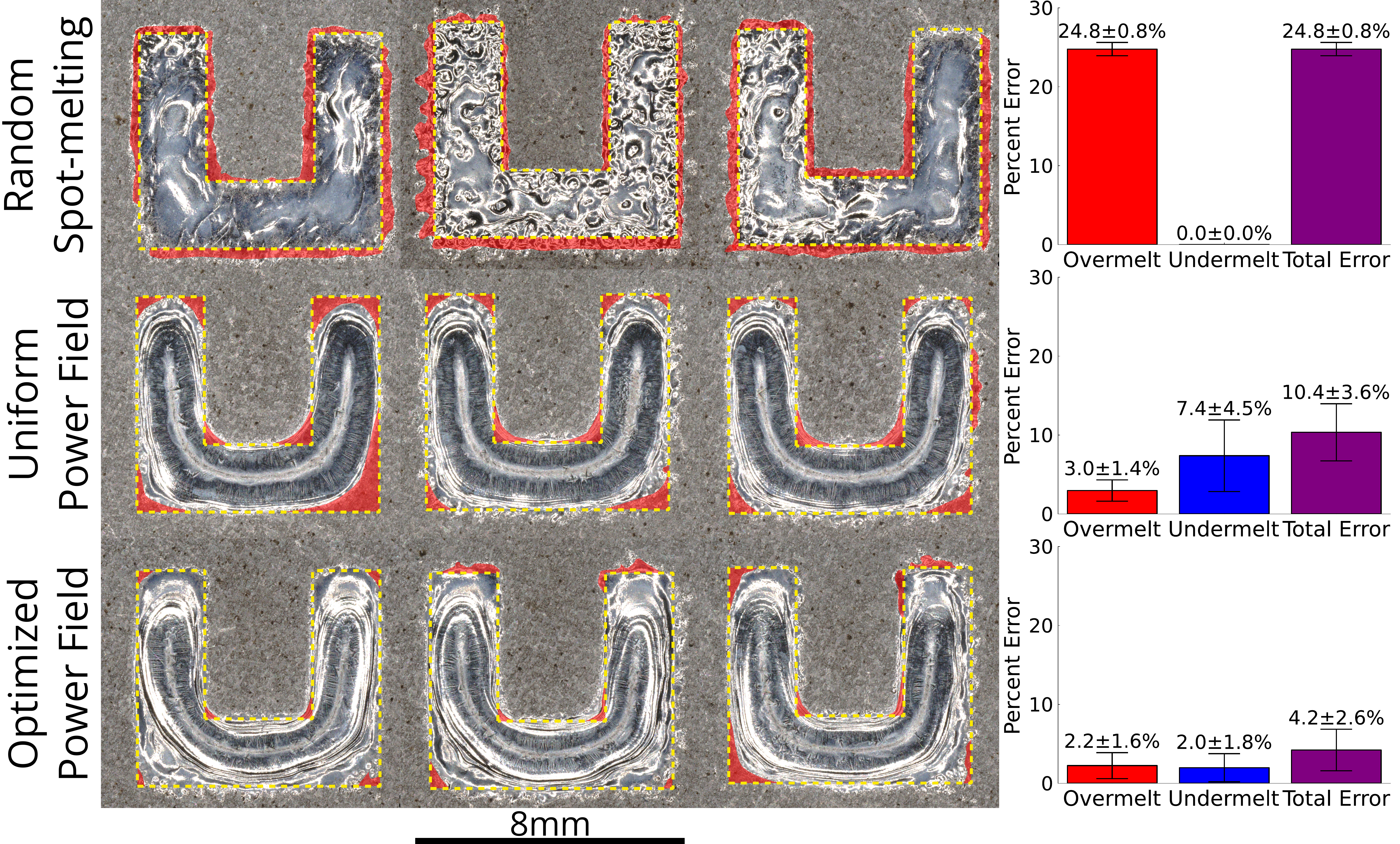}
    \caption{Three samples each printed with random spot-melting, a uniform power field, and an optimized power field. The target geometry is traced in yellow, and areas of over or under melting are highlighted in red. The mean over-, under-, and cumulative erroneous melting (sum of over- and undermelting) are shown in bar charts to the right of the cases case with 95\% confidence intervals. The experiments show a significant reduction in erroneous melting between the random and uniform cases and the uniform and optimized cases.}
    \label{fig:experimental}
\end{figure}

The resulting patches are shown in Figure \ref{fig:experimental}. As can be seen, the random spot-melting samples show a relatively sharp geometry, but with consistent over-melting / spatter, and a pitted surface. Each of the three random spot-melting samples also appears quite different, suggesting a sensitivity to initial substrate temperature conditions. 

The uniform power field samples show significant deviations in the shape, with under-melting in the outer corners and over-melting in the inner corners, and an overall pronounced curve. This is likely the result of both the differences in thermal flux between these regions as well as resultant superheat in the core of the geometry giving surface tension effects more time to bead up the molten area.

Finally, the optimized samples show both smooth surfaces and good geometric accuracy. Corner over- or under-melting is significantly reduced compared to the uniform power field, as the optimizer modulates the power field to compensate for conduction while ensuring total melting of only the target geometry while minimizing thermal variance. However, while the surface has less random pitting than the random spot-melt samples, it is certainly not free of fluid effects, with visible beading and liquid metal motion evident in the micrographs.

\section{Discussion}
The numerical results show that the optimal control approach herein presented can minimize thermal variance over a target geometry in EB-PBF by first computing an ideal optimal power-field trajectory and then greedily approximating this field with the moving beam. The experimental results show that these process plans are able to produce layers with high geometric accuracy and reduced pitting in real conditions. 

However, the experimental results also show that there remains additional work before this method can be directly deployed in the manufacture of parts. This is the result of two choices made in the approach, which the authors are considering in their ongoing work. First, since the model used is fundamentally conduction-mode, it does not model the surface tension effects that become important once the target material melts, and as such does not account for the resultant advection and bead up. This could be resolved by incorporating a mass transport model that describes advection and a term in the objective function to minimize material movement. 

Second, the current constraints require the entire target geometry to be molten simultaneously at the end of a given build step. While creating a linear, convex inequality constraint, this ensures that bulk fluid effects will play a role in final part geometry. A more relaxed constraint, such as requiring eventual, rather than simultaneous, melting could ameliorate this issue, but may introduce non-convexity into the problem, though alternative convex relaxations may be possible. 

Beyond demonstrating a new capability in EB-PBF, we believe this work serves as a demonstration for a new class of approaches to process planning and control in AM in general. The core approach of modeling quantities of interest over an AM part in state-space and solving the resultant optimal control problem can generalize to other processes and quantities. Beyond merely setting objectives directly on temperature distributions, with appropriate dynamical models it ought to be possible to pose optimal control problems in terms of microstructural and property distributions over a part, and the authors are actively engaged in demonstrating this possibility.

\section{Conclusions}
We have formulated a generalizable, modular approach to full-part process optimization for melting-based metal AM processes by the application of optimal control through the use of direct collocation. We have demonstrated the practical utility of this approach by applying it to the specific problem of ensuring thermal uniformity in the EB-PBF AM process. In this case, we have shown that the optimal control formulation results in a problem which can be efficiently solved with general purpose optimization tools to give a globally optimal process plan. These optimized process plans were then validated both in numerical simulation (showing 86\% reductions in cumulative variance) and actual experiments (showing successful printing of the geometry with a smooth surface). 

The limitations of extant AM process control and planning literature has led to conclusions that control theoretic approaches can only satisfy a single control objective, can only modulate a single parameter, and are purely reactive \cite{am_clc_ml_review}. However, such a conclusion results from the assumption that the extent of controls work in metal AM so far represents the maximum achievable limits for control and planning systems. The robotics, aerospace, and chemical industries all work with highly dynamic systems which must achieve high performance while satisfying a variety of constraints, have a large number of input variables, and often require preemptive planning. These same fields have found great success in meeting these challenges using optimal control, from running robots \cite{QuadrupedNMPC}, to self-landing rockets \cite{LosslessLanding}, to continually optimized chemical processes \cite{BIEGLER20071043}. The framework demonstrated herein serves as proof of the applicability and power of optimal control in additive manufacturing.

As such, we believe this work opens the possibility of bringing the powerful tools of optimal control theory to bear on the critical problems faced in AM process planning. With further developments in this new field -- modeling of property/microstructural dynamics to enable functional grading, advances in part discretization to allow for larger parts, improvements in specialized solvers to enable faster, potentially real-time optimization -- we hope the potential of additive techniques will be fully unlocked. 

\section*{Acknowledgements}
The authors would like to thank Alexander Myers for his input in discussions related to experiments on the EB-PBF machine.

This material is based upon work supported by the National Science Foundation Graduate Research Fellowship under Grant No. 2140739. A portion of this research was supported by the Manufacturing Futures Institute. Experimental work was performed with the support of the Next Manufacturing Center.

A portion of this research was sponsored by the Army Research Laboratory and was accomplished under Cooperative Agreement Number W911NF-20-2-0175. The views and conclusions contained in this document are those of the authors and should not be interpreted as representing the official policies, either expressed or implied, of the Army Research Laboratory or the U.S. Government. The U.S. Government is authorized to reproduce and distribute reprints for Government purposes notwithstanding any copyright notation herein. 

\section*{Preprint License}
© 2024. This manuscript version is made available under the CC-BY-NC-ND 4.0 license https://creativecommons.org/licenses/by-nc-nd/4.0/


\bibliographystyle{elsarticle-num} 
\bibliography{bib}

\begin{thebibliography}{10}
\expandafter\ifx\csname url\endcsname\relax
  \def\url#1{\texttt{#1}}\fi
\expandafter\ifx\csname urlprefix\endcsname\relax\def\urlprefix{URL }\fi
\expandafter\ifx\csname href\endcsname\relax
  \def\href#1#2{#2} \def\path#1{#1}\fi

\bibitem{XU2022112065}
Y.~Xu, B.~Mishra, S.~P. Narra, Experimental investigation of in-situ microstructural transformations in wire arc additively manufactured maraging 250-grade steel, Materials Characterization 190 (2022) 112065.
\newblock \href {https://doi.org/10.1016/j.matchar.2022.112065} {\path{doi:10.1016/j.matchar.2022.112065}}.

\bibitem{FosterBeese2017}
B.~K. Foster, A.~M. Beese, J.~S. Keist, E.~T. McHale, T.~A. Palmer, Impact of interlayer dwell time on microstructure and mechanical properties of nickel and titanium alloys, Metallurgical and Materials Transactions A 48~(9) (2017) 4411–4422.
\newblock \href {https://doi.org/10.1007/s11661-017-4164-0} {\path{doi:10.1007/s11661-017-4164-0}}.

\bibitem{DEBROY2018112}
T.~DebRoy, H.~Wei, J.~Zuback, T.~Mukherjee, J.~Elmer, J.~Milewski, A.~Beese, A.~Wilson-Heid, A.~De, W.~Zhang, Additive manufacturing of metallic components – process, structure and properties, Progress in Materials Science 92 (2018) 112--224.
\newblock \href {https://doi.org/10.1016/j.pmatsci.2017.10.001} {\path{doi:10.1016/j.pmatsci.2017.10.001}}.

\bibitem{SCHWALBACH2022143853}
E.~J. Schwalbach, J.~T. Benzing, V.~Sinha, T.~M. Butler, A.~L. Pilchak, K.~J. Chaput, N.~D. Schehl, R.~John, N.~Hrabe, Effects of local processing parameters on microstructure, texture, and mechanical properties of electron beam powder bed fusion manufactured ti–6al–4v, Materials Science and Engineering: A 855 (2022) 143853.
\newblock \href {https://doi.org/10.1016/j.msea.2022.143853} {\path{doi:10.1016/j.msea.2022.143853}}.

\bibitem{ProcessMaps}
J.~Beuth, J.~Fox, J.~Gockel, C.~Montgomery, R.~Yang, H.~Qiao, E.~Soylemez, P.~Reeseewatt, A.~Anvari, S.~Narra, N.~Klingbeil, Process mapping for qualification across multiple direct metal additive manufacturing processes, in: Proceedings of the 2013 Solid Freeform Fabrication Symposium, University of Texas at Austin, 2013.
\newblock \href {https://doi.org/10.26153/TSW/15590} {\path{doi:10.26153/TSW/15590}}.

\bibitem{BeuthPatent}
J.~L. Beuth, Process mapping of melt pool geometry (U.S. Patent 9933255B2, Apr. 2018).

\bibitem{ZHANG2021102018}
B.~Zhang, R.~Seede, L.~Xue, K.~C. Atli, C.~Zhang, A.~Whitt, I.~Karaman, R.~Arroyave, A.~Elwany, An efficient framework for printability assessment in laser powder bed fusion metal additive manufacturing, Additive Manufacturing 46 (2021) 102018.
\newblock \href {https://doi.org/10.1016/j.addma.2021.102018} {\path{doi:10.1016/j.addma.2021.102018}}.

\bibitem{NARRA2018160}
S.~P. Narra, R.~Cunningham, J.~Beuth, A.~D. Rollett, Location specific solidification microstructure control in electron beam melting of {Ti-6Al-4V}, Additive Manufacturing 19 (2018) 160--166.
\newblock \href {https://doi.org/10.1016/j.addma.2017.10.003} {\path{doi:10.1016/j.addma.2017.10.003}}.

\bibitem{FRIEDENTEMPLETON2024119632}
W.~{Frieden Templeton}, S.~Hinnebusch, S.~T. Strayer, A.~C. To, P.~C. Pistorius, S.~P. Narra, A mechanistic explanation of shrinkage porosity in laser powder bed fusion additive manufacturing, Acta Materialia 266 (2024) 119632.
\newblock \href {https://doi.org/10.1016/j.actamat.2023.119632} {\path{doi:10.1016/j.actamat.2023.119632}}.

\bibitem{freemelt}
\href{https://freemelt.com/}{freemelt} (2024).
\newline\urlprefix\url{https://freemelt.com/}

\bibitem{babu_spot}
R.~R. Dehoff, M.~M. Kirka, W.~J. Sames, H.~Bilheux, A.~S. Tremsin, L.~E. Lowe, S.~S. Babu, Site specific control of crystallographic grain orientation through electron beam additive manufacturing, Materials Science and Technology 31~(8) (2015) 931--938.
\newblock \href {https://doi.org/10.1179/1743284714Y.0000000734} {\path{doi:10.1179/1743284714Y.0000000734}}.

\bibitem{seurat}
\href{https://www.seurat.com/}{{Seurat Technologies | Metal Additive Manufacturing}} (2024).
\newline\urlprefix\url{https://www.seurat.com/}

\bibitem{vulcanforms}
\href{https://www.vulcanforms.com/}{{VulcanForms | Digital Manufacturing at Industrial Scale}} (2024).
\newline\urlprefix\url{https://www.vulcanforms.com/}

\bibitem{Gibson2020}
B.~T. Gibson, Y.~K. Bandari, B.~S. Richardson, W.~C. Henry, E.~J. Vetland, T.~W. Sundermann, L.~J. Love, Melt pool size control through multiple closed-loop modalities in laser-wire directed energy deposition of {Ti-6Al-4V}, Additive Manufacturing 32 (2020) 100993.
\newblock \href {https://doi.org/10.1016/J.ADDMA.2019.100993} {\path{doi:10.1016/J.ADDMA.2019.100993}}.

\bibitem{Li2018}
F.~Li, S.~Chen, Z.~Wu, Z.~Yan, Adaptive process control of wire and arc additive manufacturing for fabricating complex-shaped components, The International Journal of Advanced Manufacturing Technology 2018 96:1 96 (2018) 871--879.
\newblock \href {https://doi.org/10.1007/S00170-018-1590-0} {\path{doi:10.1007/S00170-018-1590-0}}.

\bibitem{Li2021}
Y.~Li, X.~Li, G.~Zhang, I.~Horváth, Q.~Han, Interlayer closed-loop control of forming geometries for wire and arc additive manufacturing based on fuzzy-logic inference, Journal of Manufacturing Processes 63 (2021) 35--47.
\newblock \href {https://doi.org/10.1016/J.JMAPRO.2020.04.009} {\path{doi:10.1016/J.JMAPRO.2020.04.009}}.

\bibitem{Tang2021}
S.~Tang, G.~Wang, H.~Song, R.~Li, H.~Zhang, A novel method of bead modeling and control for wire and arc additive manufacturing, Rapid Prototyping Journal 27 (2021) 311--320.
\newblock \href {https://doi.org/10.1108/RPJ-05-2020-0097} {\path{doi:10.1108/RPJ-05-2020-0097}}.

\bibitem{Wang2021}
Y.~Wang, J.~Lu, Z.~Zhao, W.~Deng, J.~Han, L.~Bai, X.~Yang, J.~Yao, Active disturbance rejection control of layer width in wire arc additive manufacturing based on deep learning, Journal of Manufacturing Processes 67 (2021) 364--375.
\newblock \href {https://doi.org/10.1016/j.jmapro.2021.05.005} {\path{doi:10.1016/j.jmapro.2021.05.005}}.

\bibitem{Xia2020}
C.~Xia, Z.~Pan, S.~Zhang, H.~Li, Y.~Xu, S.~Chen, Model-free adaptive iterative learning control of melt pool width in wire arc additive manufacturing, The International Journal of Advanced Manufacturing Technology 110 (2020) 2131--2142.
\newblock \href {https://doi.org/10.1007/s00170-020-05998-0/Published} {\path{doi:10.1007/s00170-020-05998-0/Published}}.

\bibitem{CRAEGHS2010505}
T.~Craeghs, F.~Bechmann, S.~Berumen, J.-P. Kruth, Feedback control of layerwise laser melting using optical sensors, Physics Procedia 5 (2010) 505--514, laser Assisted Net Shape Engineering 6, Proceedings of the LANE 2010, Part 2.
\newblock \href {https://doi.org/10.1016/j.phpro.2010.08.078} {\path{doi:10.1016/j.phpro.2010.08.078}}.

\bibitem{Shkoruta2021}
A.~Shkoruta, S.~Mishra, S.~J. Rock, Real-time image-based feedback control of laser powder bed fusion, ASME Letters in Dynamic Systems and Control 2~(2) (Jul. 2021).
\newblock \href {https://doi.org/10.1115/1.4051588} {\path{doi:10.1115/1.4051588}}.

\bibitem{WANG2023103449}
R.~Wang, B.~Standfield, C.~Dou, A.~C. Law, Z.~J. Kong, Real-time process monitoring and closed-loop control on laser power via a customized laser powder bed fusion platform, Additive Manufacturing 66 (2023) 103449.
\newblock \href {https://doi.org/10.1016/j.addma.2023.103449} {\path{doi:10.1016/j.addma.2023.103449}}.

\bibitem{WANG2020100985}
Q.~Wang, P.~P. Michaleris, A.~R. Nassar, J.~E. Irwin, Y.~Ren, C.~B. Stutzman, Model-based feedforward control of laser powder bed fusion additive manufacturing, Additive Manufacturing 31 (2020) 100985.
\newblock \href {https://doi.org/10.1016/j.addma.2019.100985} {\path{doi:10.1016/j.addma.2019.100985}}.

\bibitem{DRUZGALSKI2020101169}
C.~Druzgalski, A.~Ashby, G.~Guss, W.~King, T.~Roehling, M.~Matthews, Process optimization of complex geometries using feed forward control for laser powder bed fusion additive manufacturing, Additive Manufacturing 34 (2020) 101169.
\newblock \href {https://doi.org/10.1016/j.addma.2020.101169} {\path{doi:10.1016/j.addma.2020.101169}}.

\bibitem{Forslund2021}
R.~Forslund, A.~Snis, S.~Larsson, A greedy algorithm for optimal heating in powder-bed-based additive manufacturing, Journal of Mathematics in Industry 11~(1) (8 2021).
\newblock \href {https://doi.org/10.1186/s13362-021-00110-x} {\path{doi:10.1186/s13362-021-00110-x}}.

\bibitem{Xiong2022}
J.~Xiong, H.~Chen, S.~Zheng, G.~Zhang, Feedback control of variable width in gas metal arc-based additive manufacturing, Journal of Manufacturing Processes 76 (2022) 11--20.
\newblock \href {https://doi.org/10.1016/J.JMAPRO.2022.02.008} {\path{doi:10.1016/J.JMAPRO.2022.02.008}}.

\bibitem{Shi2023}
Y.~Shi, S.~Gong, H.~Xu, G.~Yang, J.~Qiao, Z.~Wang, J.~Zhang, B.~Qi, Electron beam metal additive manufacturing: Defects formation and in-process control, Journal of Manufacturing Processes 101 (2023) 386--431.
\newblock \href {https://doi.org/10.1016/J.JMAPRO.2023.06.013} {\path{doi:10.1016/J.JMAPRO.2023.06.013}}.

\bibitem{OLLEAK2024100197}
A.~Olleak, E.~Adcock, S.~Hinnebusch, F.~Dugast, A.~D. Rollett, A.~C. To, Understanding the role of geometry and interlayer cooling time on microstructure variations in {LPBF} {Ti6Al4V} through part-scale scan-resolved thermal modeling, Additive Manufacturing Letters 9 (2024) 100197.
\newblock \href {https://doi.org/10.1016/j.addlet.2024.100197} {\path{doi:10.1016/j.addlet.2024.100197}}.

\bibitem{smartscan}
K.~S. Ramani, C.~He, Y.-L. Tsai, C.~E. Okwudire, {SmartScan}: An intelligent scanning approach for uniform thermal distribution, reduced residual stresses and deformations in {PBF} additive manufacturing, Additive Manufacturing 52 (2022) 102643.
\newblock \href {https://doi.org/10.1016/j.addma.2022.102643} {\path{doi:10.1016/j.addma.2022.102643}}.

\bibitem{Schmidt2023}
J.~Schmidt, A.~F{\"u}genschuh, Trajectory optimization for arbitrary layered geometries in wire-arc additive manufacturing, Optimization and Engineering (6 2023).
\newblock \href {https://doi.org/10.1007/s11081-023-09813-z} {\path{doi:10.1007/s11081-023-09813-z}}.

\bibitem{Karp1972}
R.~M. Karp, Reducibility among Combinatorial Problems, Springer US, 1972, p. 85–103.
\newblock \href {https://doi.org/10.1007/978-1-4684-2001-2\_9} {\path{doi:10.1007/978-1-4684-2001-2\_9}}.

\bibitem{Wood2023Theory}
N.~Wood, D.~J. Hoelzle, On the controllability and observability of temperature states in metal powder bed fusion, Journal of Dynamic Systems, Measurement, and Control 145 (3 2023).
\newblock \href {https://doi.org/10.1115/1.4056326} {\path{doi:10.1115/1.4056326}}.

\bibitem{wood2019seeing}
N.~Wood, D.~J. Hoelzle, `{Seeing}'the temperature inside the part during the powder bed fusion process, in: 2019 International Solid Freeform Fabrication Symposium, University of Texas at Austin, 2019.

\bibitem{wood2021ensemble}
N.~Wood, E.~Schwalbach, A.~Gillman, D.~J. Hoelzle, The ensemble kalman filter as a tool for estimating temperatures in the powder bed fusion process, in: 2021 American Control Conference (ACC), IEEE, 2021, pp. 4369--4375.

\bibitem{MurrayNetworks}
R.~Saber, R.~Murray, Consensus protocols for networks of dynamic agents, in: Proceedings of the 2003 American Control Conference, 2003., Vol.~2, 2003, pp. 951--956.
\newblock \href {https://doi.org/10.1109/ACC.2003.1239709} {\path{doi:10.1109/ACC.2003.1239709}}.

\bibitem{DIRCOL}
C.~Hargraves, S.~Paris, Direct trajectory optimization using nonlinear programming and collocation, AIAA J. Guidance 10 (1987) 338--342.
\newblock \href {https://doi.org/10.2514/3.20223} {\path{doi:10.2514/3.20223}}.

\bibitem{Wächter2006}
A.~W{\"a}chter, L.~T. Biegler, On the implementation of an interior-point filter line-search algorithm for large-scale nonlinear programming, Mathematical Programming 106~(1) (2006) 25--57.
\newblock \href {https://doi.org/10.1007/s10107-004-0559-y} {\path{doi:10.1007/s10107-004-0559-y}}.

\bibitem{HSL_MA97}
J.~S. JD~Hogg, {HSL MA97}: a bit-compatible multifrontal code for sparse symmetric systems, Tech. rep., Rutherford Appleton Laboratory (2011).

\bibitem{symbolics_jl}
S.~Gowda, Y.~Ma, A.~Cheli, M.~Gw\'{o}\'{z}zd\'{z}, V.~B. Shah, A.~Edelman, C.~Rackauckas, High-performance symbolic-numerics via multiple dispatch, ACM Commun. Comput. Algebra 55~(3) (2022) 92–96.
\newblock \href {https://doi.org/10.1145/3511528.3511535} {\path{doi:10.1145/3511528.3511535}}.

\bibitem{RevelsLubinPapamarkou2016}
J.~{Revels}, M.~{Lubin}, T.~{Papamarkou}, Forward-mode automatic differentiation in {J}ulia, arXiv:1607.07892 [cs.MS] (2016).

\bibitem{SparseDiffTools}
JuliaDiff, {SparseDiffTools.jl}, \url{github.com/JuliaDiff/SparseDiffTools.jl} (2024).

\bibitem{mills}
K.~Mills, Recommended values of thermophysical properties for selected commercial alloys, Woodhead Publishing, 2002, pp. 135--142.

\bibitem{POTRA2000281}
F.~A. Potra, S.~J. Wright, Interior-point methods, Journal of Computational and Applied Mathematics 124~(1) (2000) 281--302, numerical Analysis 2000. Vol. IV: Optimization and Nonlinear Equations.
\newblock \href {https://doi.org/10.1016/S0377-0427(00)00433-7} {\path{doi:10.1016/S0377-0427(00)00433-7}}.

\bibitem{am_clc_ml_review}
D.~Gunasegaram, A.~Barnard, M.~Matthews, B.~Jared, A.~Andreaco, K.~Bartsch, A.~Murphy, Machine learning-assisted in-situ adaptive strategies for the control of defects and anomalies in metal additive manufacturing, Additive Manufacturing 81 (2024) 104013.
\newblock \href {https://doi.org/10.1016/j.addma.2024.104013} {\path{doi:10.1016/j.addma.2024.104013}}.

\bibitem{QuadrupedNMPC}
M.~Neunert, M.~Stäuble, M.~Giftthaler, C.~D. Bellicoso, J.~Carius, C.~Gehring, M.~Hutter, J.~Buchli, Whole-body nonlinear model predictive control through contacts for quadrupeds, IEEE Robotics and Automation Letters 3~(3) (2018) 1458--1465.
\newblock \href {https://doi.org/10.1109/LRA.2018.2800124} {\path{doi:10.1109/LRA.2018.2800124}}.

\bibitem{LosslessLanding}
B.~Açıkmeşe, J.~M. Carson, L.~Blackmore, Lossless convexification of nonconvex control bound and pointing constraints of the soft landing optimal control problem, IEEE Transactions on Control Systems Technology 21~(6) (2013) 2104--2113.
\newblock \href {https://doi.org/10.1109/TCST.2012.2237346} {\path{doi:10.1109/TCST.2012.2237346}}.

\bibitem{BIEGLER20071043}
L.~T. Biegler, An overview of simultaneous strategies for dynamic optimization, Chemical Engineering and Processing: Process Intensification 46~(11) (2007) 1043--1053, special Issue on Process Optimization and Control in Chemical Engineering and Processing.
\newblock \href {https://doi.org/10.1016/j.cep.2006.06.021} {\path{doi:10.1016/j.cep.2006.06.021}}.

\end{thebibliography}

\end{document}